# Portfolio Optimization under Habit Formation

Roman Naryshkin[1] and Matt Davison[1,2]

[1]*Department of Applied Mathematics, University of Western Ontario*

[2]*Department of Statistical and Actuarial Sciences, University of Western Ontario*

February 20, 2013

**Abstract**

The "standard" Merton formulation of optimal investment and consumption involves optimizing the integrated lifetime utility of consumption, suitably discounted, together with the discounted future bequest. In this formulation the utility of consumption at any given time depends only on the amount consumed at that time. However, it is both theoretically and empirically reasonable that an individuals utility of consumption would depend on past consumption history. Economists term this "Habit Formation". We introduce a new formulation of habit formation which allows non-addictive consumption patterns for a wide variety of utility specification. In this paper we construct a simple mathematical description of this habit formation and present numerical solutions. We compare the results with the standard ones and draw insights obtained from the habit formation. The consumption path tends to increase with time and be less sensitive to the market fluctuations, which perfectly reflects the existence of habit persistence of an investor. At the same time, his decreasing risk aversion, which seems to be in contradiction with the empirical evidence, can be explained within the limitations of the model.

## 1 Introduction

The original portfolio selection theory, originally proposed by Markowitz [1], was a one-period model maximizing the expected terminal wealth of investments on the basis of mean and variance. The modern paradigm considers a many-period model with a hedging strategy, where the investor's utility depends on the consumption of goods. The principles and the main results of the model were given in the seminal works by Merton [2] for the continuous time case and Samuelson [3] for the discrete time case and form the *standard* well-developed portfolio optimization theory [4].

In the standard theory, the investor's utility $U$ of consumption $C$ at any given time $t$ depends only on the amount consumed at that time $U[C(t)]$ and doesn't include any information about the past consumption or constantly increasing demands and/or needs when the investor becomes accustomed to a new standard of living. These facts have continuously been under critique, e.g. [5] introduces a theory where the utility is gained from the changes of the variable, not from its current absolute value, and [6] discusses psychological aspects, arguing that decisions are made using relative comparisons, not absolute. Another drawback of the standard theory is that it leads to some paradoxical results. The most famous one is the so-called "Equity Premium Puzzle" [7], whose resolution was found within the class of habit formation models [8]. Another "Merton Paradox" was discussed in [9], concluding that, in contrast to the solution, the risk aversion of the young should differ from that of the old, in general changing with the investor's age.

One way to resolve the problems is to account for intolerance for a decline in standard of living by a non-decreasing consumption constraint $C' \geqslant 0$ in continuous time [10] and in discrete time [11]. This can be treated as an extreme case where a utility function is unbounded from below for $C' < 0$, which is too restrictive and may lead to a bankruptcy of the investor before the time horizon of the model.

Another resolution of the paradoxes was found in the models with *Habit Formation* where the utility of consumption depends not only on the current consumption rate $C(t)$ but also on the individuals past consumption history $Z(t)$[1]. The general properties of the deterministic model with Habit Formation, where $U = U[C(t), Z(t)]$ were given in [12]. The closed-form solutions of a stochastic model were found only for a utility function of the very special "additive" form $U = U[C(t) - Z(t)]$ in continuous time, where $U$ is either a power or an exponential function [13], [8] (infinite time horizon), and [14], [15] (finite time horizon). The last word in this direction are [16] and [17], where it is admitted that there are no known cases beyond the additive

[1]The typical assumption is $Z(t) = Z_0 e^{-at} + A\int_0^t e^{-a(t-\tau)} C(\tau) d\tau$, but it is more intuitive to consider $Z(t) = \frac{1}{t}\int_0^t C(\tau) d\tau$

one $U[C(t)-Z(t)]$ that allows for a closed-form solution.

Even though some nice analytical solutions are obtained in both above described approaches, they seem to be restrictive enough ($C' \geqslant 0$ in the first approach and $C(t) \geqslant Z(t)$ in the second) to lead to a "premature bankruptcy" in cases where the wealth randomly drops below some parameter-dependent level (see e.g. [13]). In short, the existing Habit Formation models can be described by the relation "additive=addictive", since the minimum allowed consumption level is a non-decreasing function of time $C_{\min}(t+\Delta t) \geqslant C_{\min}(t)$. Therefore, we need a model that has a better hedging strategy, and the known analytical solutions for Habit Formation seem not to be greatly important in practice.

The present paper considers a model that is free from the above mentioned paradoxes at the cost of not having any closed-form analytical solutions. However, numerical solutions can be obtained allowing practical asset allocation decisions. Another advantage of the model is that it accepts arbitrary utility functions. This allows more accurate description of investor's preferences and risk aversion in contrast to the available analytical solutions that exist only for power and exponential utility functions.

## 2 The Habit Formation Utility Function

A Habit Formation utility function that has no constraints on the current consumption $C(t)$ (i.e. investor's preferences are not addictive) is chosen to have the form

$$U[C(t),\bar{C}(t)] = U\left[\frac{C(t)}{C_0+\beta\bar{C}(t)}\right], \tag{2.1}$$

where $C_0$ is the inherited level of consumption at the moment $t=0$, positive constant $\beta$ is the memory parameter that gives an estimate of the influence of the past consumption on the present one ($\beta = 0$ corresponds to the standard Merton's problem), and $\bar{C}(t)$ is the averaged past consumption until the present moment $t$ :

$$\bar{C}(t) = \frac{1}{t}\int_0^t C(\tau)d\tau. \tag{2.2}$$

The utility function (2.1) measures the relative satisfaction of the current consumption with respect to the averaged consumption in the past (in contrast to the absolute satisfaction for the additive-addictive utility function [12]–[17]).

Further we will consider the CRRA-type utility function

$$U\left[C(t),\bar{C}(t)\right]=\frac{1}{\gamma}\left(\frac{C(t)}{C_0+\beta C(t)}\right)^{\gamma}, \tag{2.3}$$

however the general dynamic programming method presented below easily allows extension of the results to an arbitrary utility function having the form (2.1).

# 3 The Two-Asset Optimization Problem

In the simplest two-asset model, consumption is financed out of a portfolio consisting a risky asset (stock) with a return rate following a lognormal random walk with drift $\mu$ and volatility $\sigma$, and a riskless asset (bond) with an interest rate $r$. The control functions are the consumption rate $C(t)$ and the risky portfolio weight $\omega(t)$, which is a fraction of wealth invested in the stock, so that the total wealth $W(t)$ satisfies the budget equation:

$$dW=-Cdt+(1-\omega)Wrdt+\omega W\left(\mu dt+\sigma\sqrt{dt}\,z\right),\quad z\in N(0,1), \tag{3.1}$$

subject to the initial condition $W(0)=W_0$.

An investor maximizes his suitably discounted expected future utility

$$J[W(0),\bar{C}(0),0]=\max_{C(t),\omega(t)}\mathrm{E}_0\left\{\int_0^T e^{-\rho t}\,U\left[\frac{C(t)}{C_0+\beta C(t)}\right]dt+B[W(T),T]\right\}, \tag{3.2}$$

where $T$ is the investor's time horizon, $\rho$ is the discount factor that reflects investor's time preference of consumption, and $B[W(T),T]$ is the bequest function. A convenient choice of the bequest function is such where the functional dependence on the terminal wealth $W(T)$ is of the same form as the utility function. We will assume that

$$B[W(T),T]=be^{-\rho T}U[W(T)] \tag{3.3}$$

with $b\geqslant 0$. The case with $b=0$ corresponds to the no-bequest formulation of the problem where an investor consumes all available wealth at the time horizon.

# 4 Numerical Algorithm

An analytical solution of the above described problem is no longer available and we have to use numerical optimization methods. For this purpose, we adopt the discrete version of the stochastic dynamic programming method.

First, we discretize time by dividing the interval $[0,T]$ into $N$ equidistant discrete points $t_i = (i-1/2)\Delta t,\ i = 1..N$ with $\Delta t = T/N$. Then, the dynamic programming algorithm leads to the backward recursive functional equation:

$$J_i[W_i,\bar{C}_i] = \max_{0\leqslant C_i \leqslant \frac{W_i}{\Delta t}} \max_{0\leqslant \omega_i \leqslant 1} \mathrm{E}_0 \left\{ e^{-\rho t_i}\, U\left[\frac{C_i}{C_0+\beta C_i}\right]\Delta t + J_{i+1}[W_{i+1},\bar{C}_{i+1}]\right\}, \tag{4.1}$$

where

$$W_{i+1} = (W_i - C_i\Delta t)\left[1+(1-\omega_i)r\Delta t + \omega_i\left(\mu\Delta t + \sigma\sqrt{\Delta t}z_i\right)\right], \tag{4.2}$$

and

$$\bar{C}_{i+1} = \frac{1}{i}\sum_{k=1}^{i} C_k = \frac{C_i}{i} + \frac{i-1}{i}\bar{C}_i. \tag{4.3}$$

The base of the recursive equation (4.1) is

$$J_{N+1}[W_{N+1},\bar{C}_{N+1}] = B[W_{N+1},T] = be^{-\rho T}U[W_{N+1}]. \tag{4.4}$$

The equation (4.1) can be solved numerically to determine the values for $J_i[W_i,\bar{C}_i]$ and then the optimal policies for the consumption $C_i^*$ and investment $\omega_i^*$.

Second, for a numerical purpose, we have to discretize the "space" variables $W$ and $\bar{C}$. This is done by putting the dynamic programming method on a mesh in those variables. However, the final result for $J_i[W,\bar{C}]$ will be continuous since it will be parabolically interpolated between the nodes giving the second order of accuracy with respect to the stepsizes of time and space variables.

Therefore, the dynamic programming is carried out in a 3-dimensional discretized space with coordinates $t,W,\bar{C}$ where, at each point, maxima over $C_i$ and $\omega_i$ should be found and the expectation over $z_i$ must be evaluated.

# 5 Results and Discussion

The numerical results are obtained using the CRRA-type utility function (2.3) with the following set of parameters: initial wealth $W_0 = \$1,000,000$, time horizon $T = 10$ years, inherited consumption level $C_0 = W_0/T$, memory parameter $\beta = 0.1$ (Weak Habit Formation) and $\beta = 1$ (Strong Habit Formation), the riskless asset interest rate $r = 3\%$, the drift and the volatility of the risky asset $\mu = 5\%$ and $\sigma = 25\%$ correspondingly, the investor's relative risk aversion $1-\gamma = 0.5$, and the discount factor takes values $\rho = 0\%$, and $\rho = 10\%$. The bequest function is assumed to be zero $b = 0$ until the subsection 5.5, where the value of $b$ is chosen to guarantee that the final wealth be equal to the initial one.

## 5.1 Merton's solution

The standard Merton's solution can be restored from the Habit Formation results by setting the memory parameter to zero $\beta = 0$. Moreover, there is an analytical solution available for this case. The optimum consumption rate reads [2]

$$C^*(t) = \frac{\nu W(t)}{1 - e^{-\nu(T-t)}}, \quad \nu = \frac{1}{1-\gamma}\left\{\rho - \gamma\left[r + \frac{(\mu - r)^2}{2\sigma^2(1-\gamma)}\right]\right\}, \tag{5.1}$$

while the optimum portfolio weight is constant

$$\omega^*(t) = \frac{\mu - r}{\sigma^2(1-\gamma)}. \tag{5.2}$$

In the following subsections the Merton's solution ($\beta = 0$) will be compared with the Weak Habit Formation solution ($\beta = 0.1$) as well as with the Strong Habit Formation solution ($\beta = 1$).

## 5.2 Weak Habit Formation

The numerical results will be presented for a particular realization of a stochastic process of a stock price as well as for the expected value of a stock price estimated as an average over 1000 different realizations (see Fig. 1).

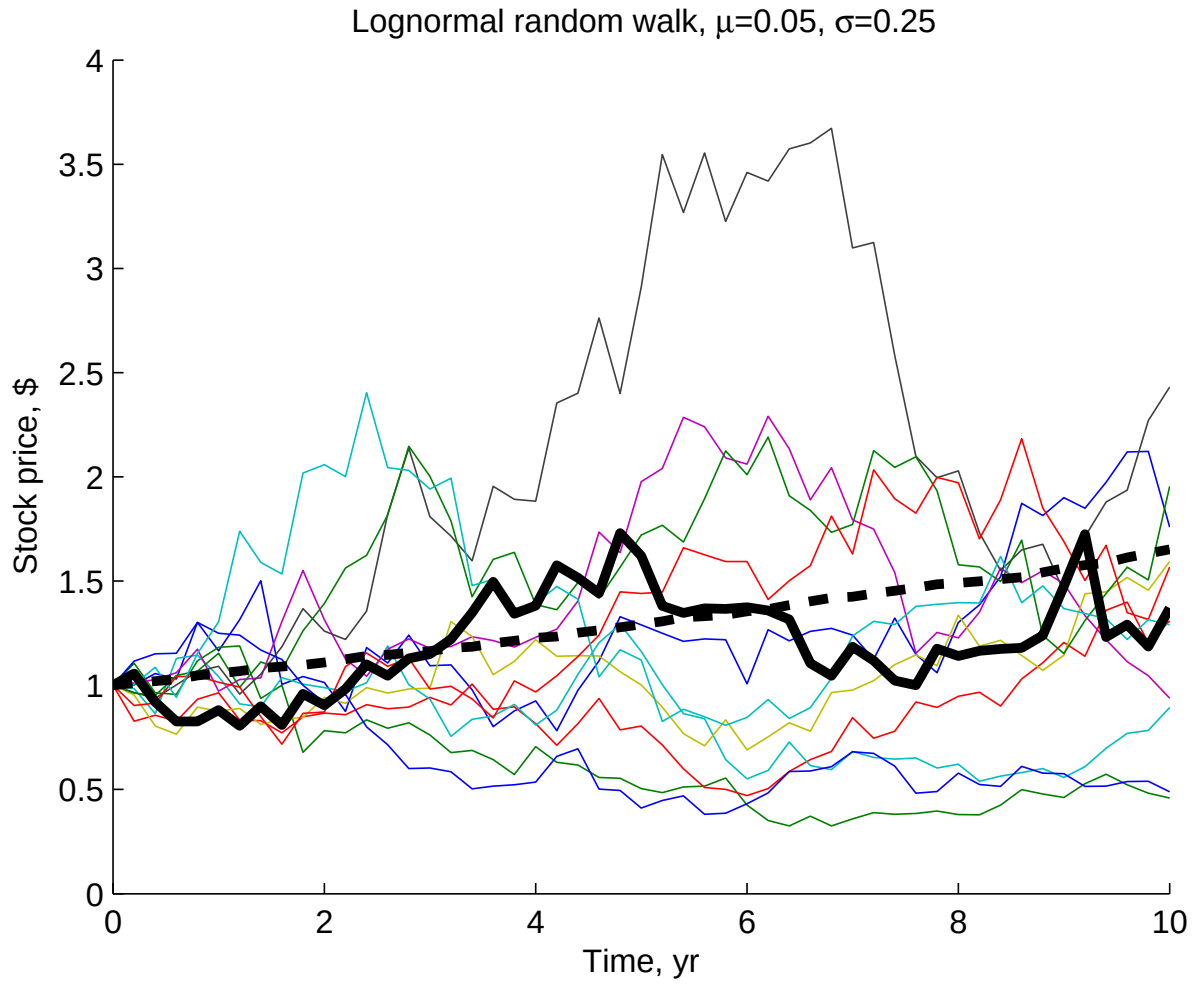

**Figure 1**. Stock price realizations and the expected price.

On Fig. 2 we show the consumption $C$ and portfolio weight $\omega$ as functions of time for the discount factor $\rho = 10\%$.

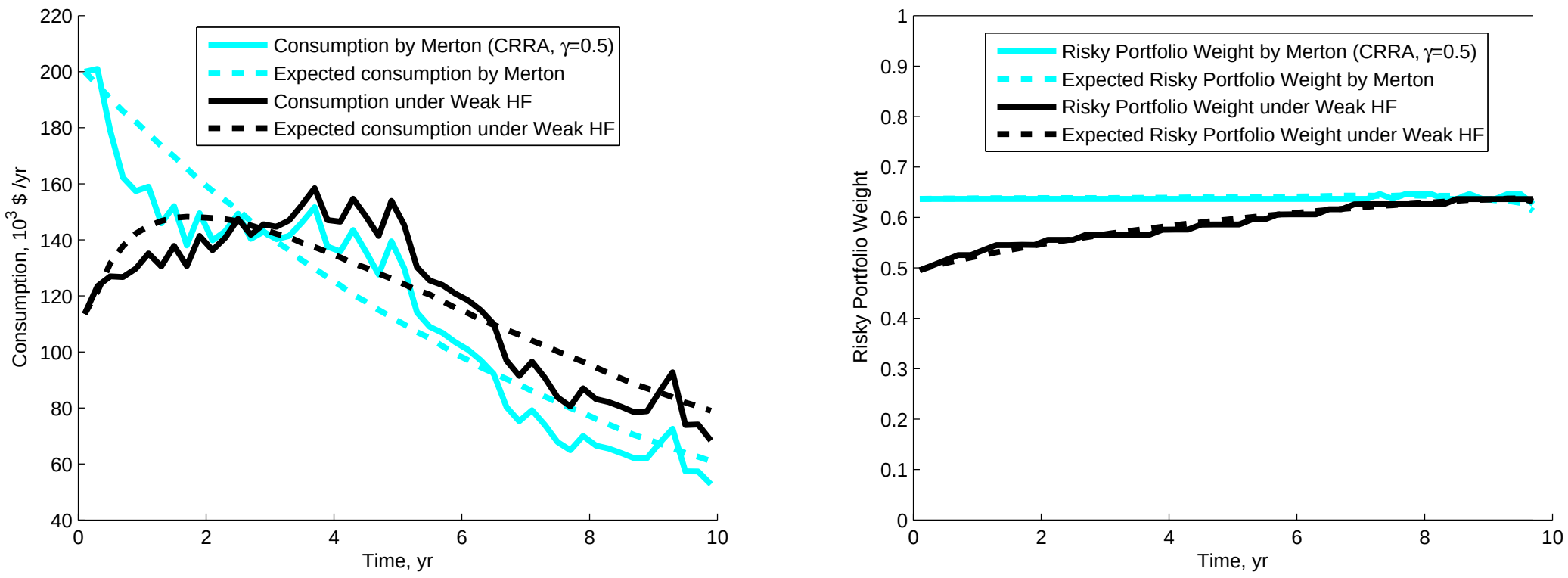

**Figure 2**. Optimum consumption $C(t)$ and risky portfolio weight $\omega(t)$ for $\rho = 10\%, \beta = 0.1$

The solid grey lines correspond to the standard Merton solution for a particular stock price realization, while the dashed grey curves correspond to the expected values of consumption and portfolio weight. The solid and dashed black lines correspond to the optimum solution of the optimization problem under Weak Habit Formation.

The consumption and portfolio weight for the trivial discount factor $\rho = 0\%$ are shown on Fig. 3.

First, as expected, the optimum consumption rate for the Habit Formation case is lower than Merton's in the beginning, giving the opportunity to an investor to have a larger consumption rate later as his habits form. In addition, the fluctuations of the market are more smoothed by the Habit Formation (in particular, see Fig. 9 with $\beta = 1$), which means that the consumption is no longer proportional to the total wealth (see subsection 5.3). More importantly, the consumption policy in our model doesn't lead to bankruptcy as invariably occurred in the case of time-additive utility function [13].

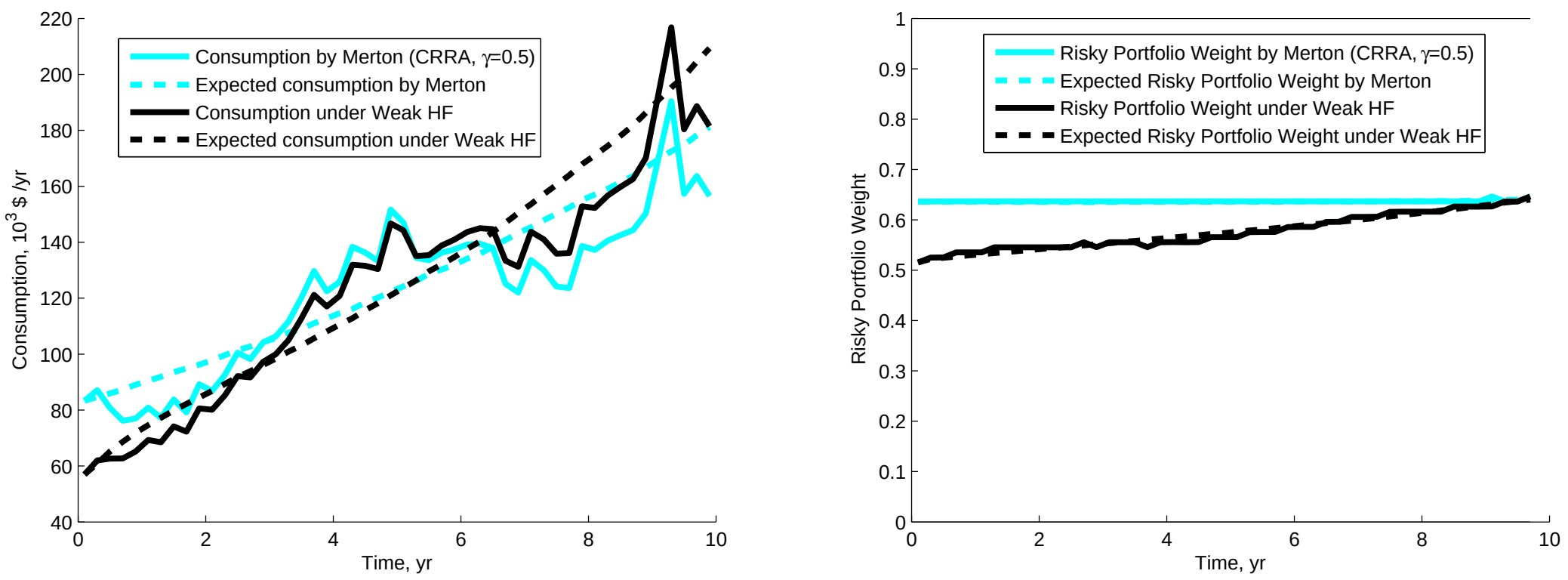

**Figure 3**. Optimum consumption $C(t)$ and risky portfolio weight $\omega(t)$ for $\rho = 0\%, \beta = 0.1$

Secondly, the portfolio weight has a somewhat surprising behavior: it shows that an investor becomes more and more risk-seeking as he ages and the fraction of wealth invested in the stock approaches to the Merton ratio only near the time horizon. An explanation for this behavior is given in section 5.6. Nevertheless, the risk aversion is always bigger than that of Merton's investor. The portfolio weight is no longer constant in time and this fact helps to avoid the famous paradox of the constant Merton solution (see [9]).

Finally, we can see the impact of the discount factor that represents the investor's time preferences and/or risk aversion against his uncertain lifetime. When the discount factor (e.g. $\rho = 0\%$) is less than the expected return of the portfolio, which is about 4%, the consumption path increases in time for both the standard and the Habit Formation solutions. Only when $\rho$ is bigger than the expected portfolio return (e.g. $\rho = 10\%$) does the qualitative difference between the consumptions become evident: Merton's path decreases due to the large time preference of an investor, while the Habit Formation path has a peak showing that the habit persistence dominates over the time preference in the beginning. Obviously, the bigger the memory parameter, the bigger the dominance over time preferences will be (see subsection 5.4). However, the port-folio weight shows only slight sensitivity with respect to the discount factor in accordance with the Merton solution where it doesn't depend on the discount factor at all.

## 5.3 Wealth-dependence of the optimum solutions

In the previous subsection, the time evolution of the optimum consumption $C(t)$ and portfolio weight $\omega(t)$ was shown. Another way of looking at the results is to get rid of the explicit dependence on time and investigate the "internal" relationships $C(W)$ and $\omega(W)$. Before doing that we notice that the wealth itself depends on the time as shown on the Fig. 4. Notice how convex shape of $W(t)$ becomes concave when the discount factor $\rho$ exceeds the expected portfolio return.

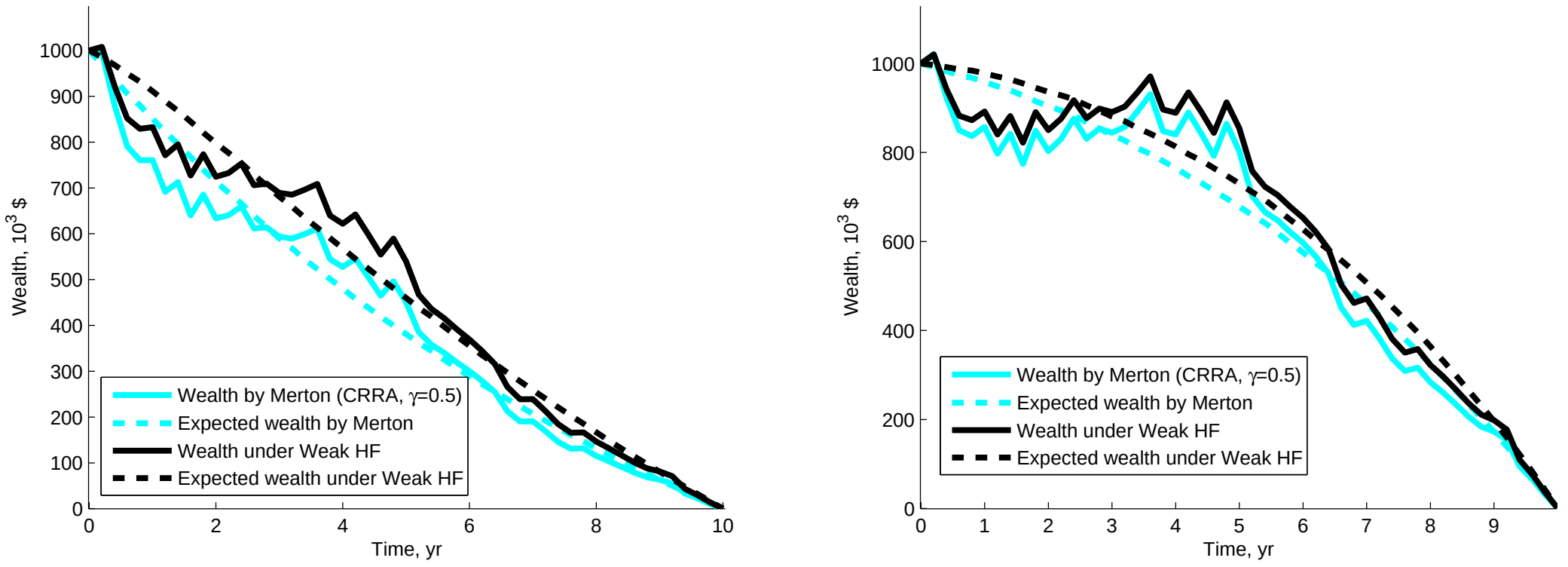

**Figure 4**. Wealth evolution $W(t)$ for $\beta = 0.1$, $\rho = 10\%$ (left) and $\rho = 0\%$ (right)

We can numerically invert the functions $W(t)$ to get $t(W)$ and plug them into $C(t)$ and $\omega(t)$. The results of such a "change of variables" are shown on Figs. 5 and 6.

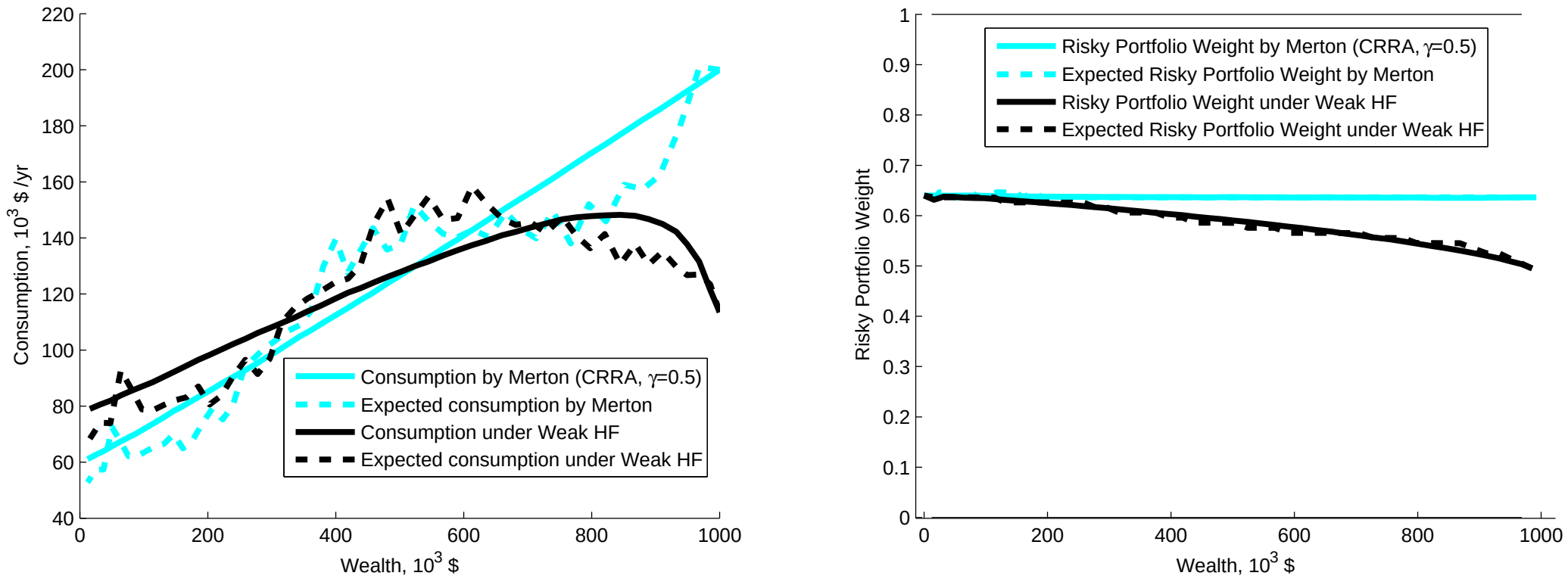

**Figure 5**. Consumption and risky portfolio weight versus expected total wealth, $\rho = 10\%, \beta = 0.1$

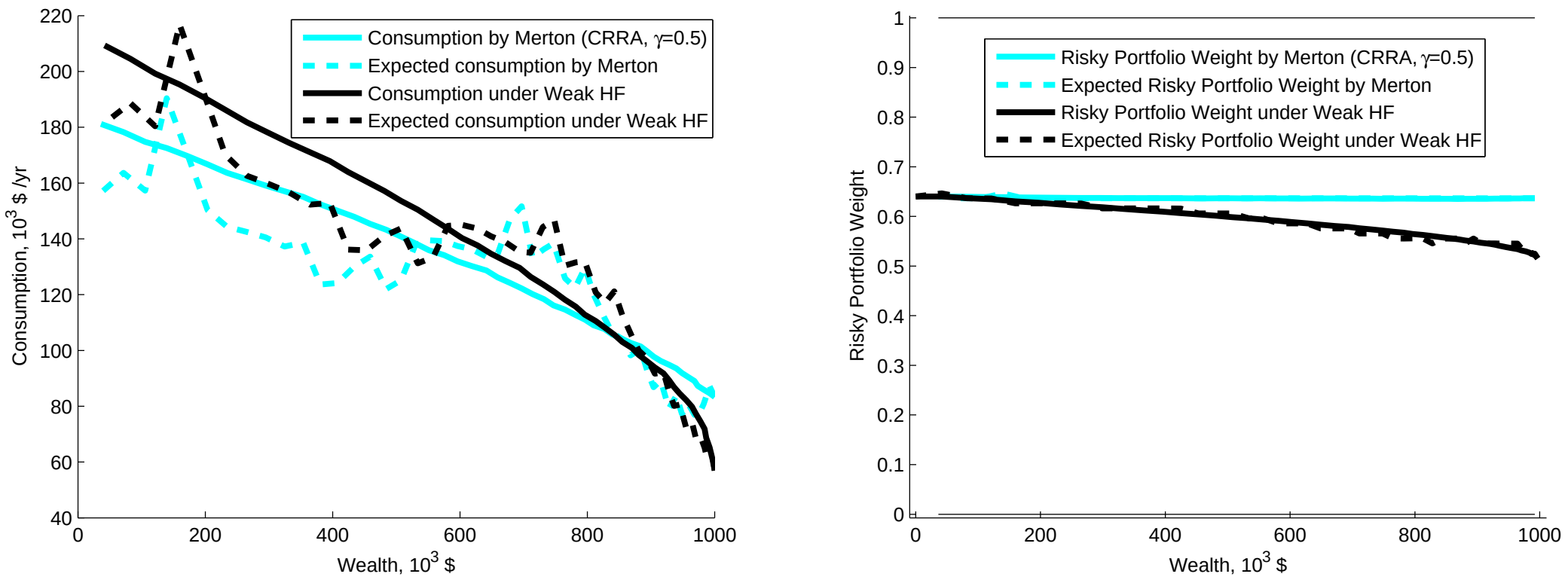

**Figure 6**. Consumption and risky portfolio weight versus expected total wealth, $\rho = 0\%, \beta = 0.1$

We can see that in the Merton's solution (grey lines), the expected consumption is proportional to the expected total wealth (with the approximate slope $0.025(\rho - 4\%)$) and the portfolio weight is independent of the total wealth. However, the Habit Formation changes linear dependencies to the convex functions of wealth for both consumption and portfolio weight.

It is interesting to note that our results for portfolio weight as a decreasing function of wealth show completely different behavior comparing to the additive utility function results [13] for habit formation, where the portfolio weight is an increasing function of time and approaches Merton's proportion at infinite large wealth. At the same time, the model [13] allows for a premature bankruptcy if the wealth drops below some critical amount (the consumption as a function of wealth has a threshold below which the consumption is undefined). Therefore, as argued in the introduction, we think that our model may better describe the real habit persistence of an investor under the same assumptions.

## 5.4 Strong Habit Formation

It is also interesting to vary the memory parameter $\beta$ that measures the habit persistence of an investor. We can expect that the bigger the $\beta$, the bigger the deflection from the Merton's solution will be. This is indeed true and the numerical results for the Strong Habit Formation with memory parameter $\beta = 1$ (black line) are compared with the results for Weak Habit Formation with $\beta = 0.1$ (dash-and-dot curve) on the figures below.

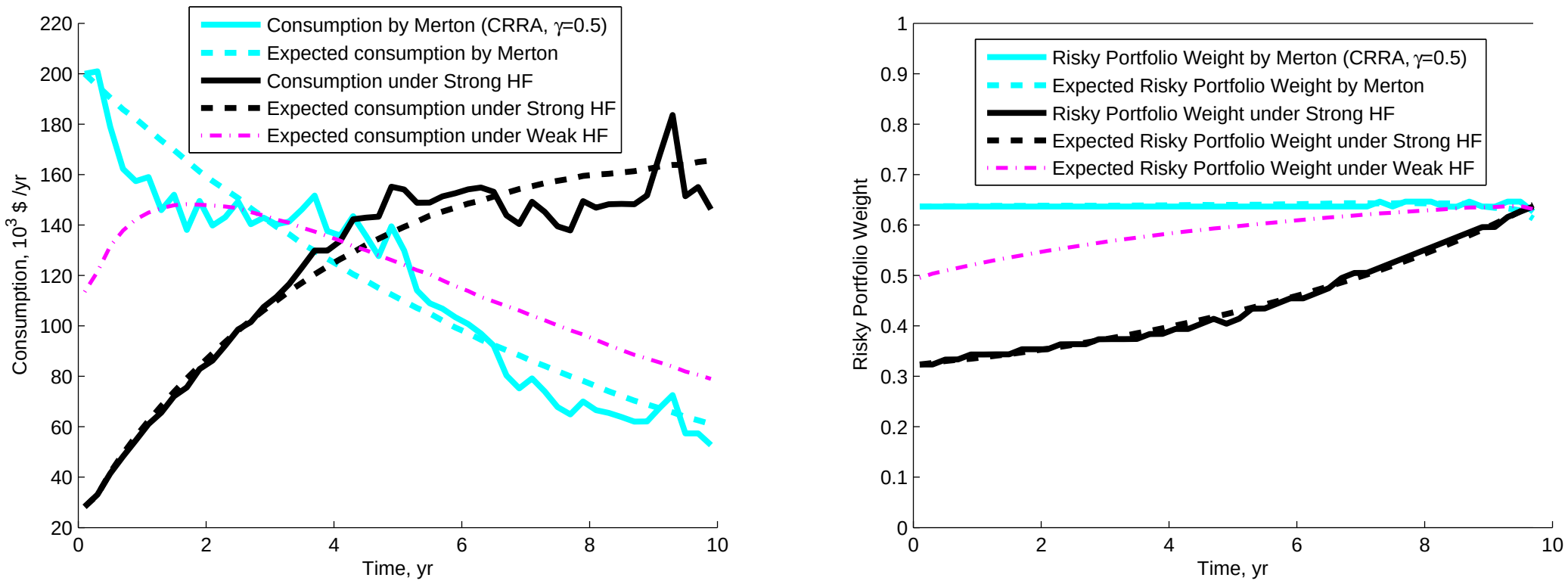

**Figure 7**. Consumption and risky portfolio weight under Strong Habit Formation $\beta = 1.0$ with $\rho = 10\%$

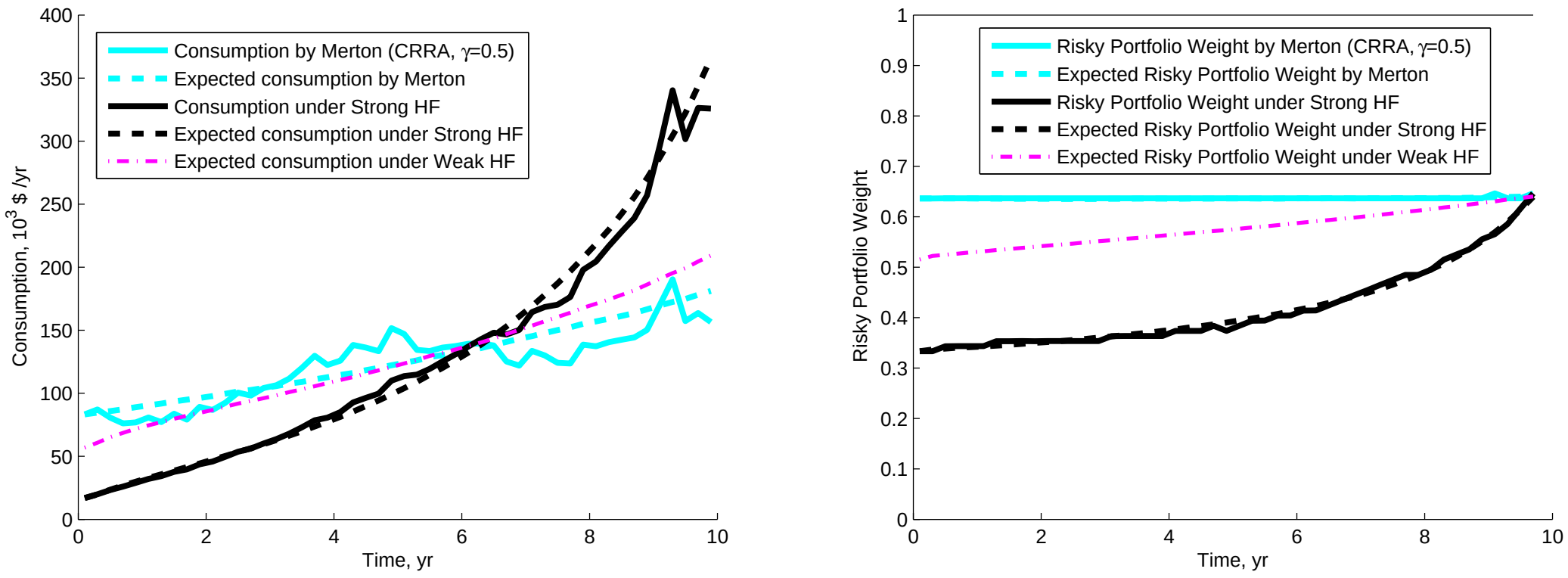

**Figure 8**. Consumption and risky portfolio weight under Strong Habit Formation $\beta = 1.0$ with $\rho = 0\%$

We see that as $\beta$ becomes bigger, the consumption path becomes more and more convex, and its maximum slowly moves to later times. Also, the initial consumption becomes smaller which means that the habit formation of an investor becomes increasingly dominant over his time preferences. On the other hand, the portfolio selection results show that the bigger the investor's habit persistence, the more risk averse he will be in the early stages of his investment. And under no circumstances will the investor be less risk averse than Merton's investor so that the Merton ratio is never exceeded.

In addition, as we mentioned before, the sensitivity of consumption rate to market fluctuations decreases as the memory parameter increases (see Fig. 9 for comparison between $\beta = 0$ and $\beta = 1$), while the portfolio weight seems to be entirely insensitive, which might be an indication that the portfolio weight is independent of wealth and depends only on time (and therefore the dependence of $\omega$ on wealth is completely implicit through time).

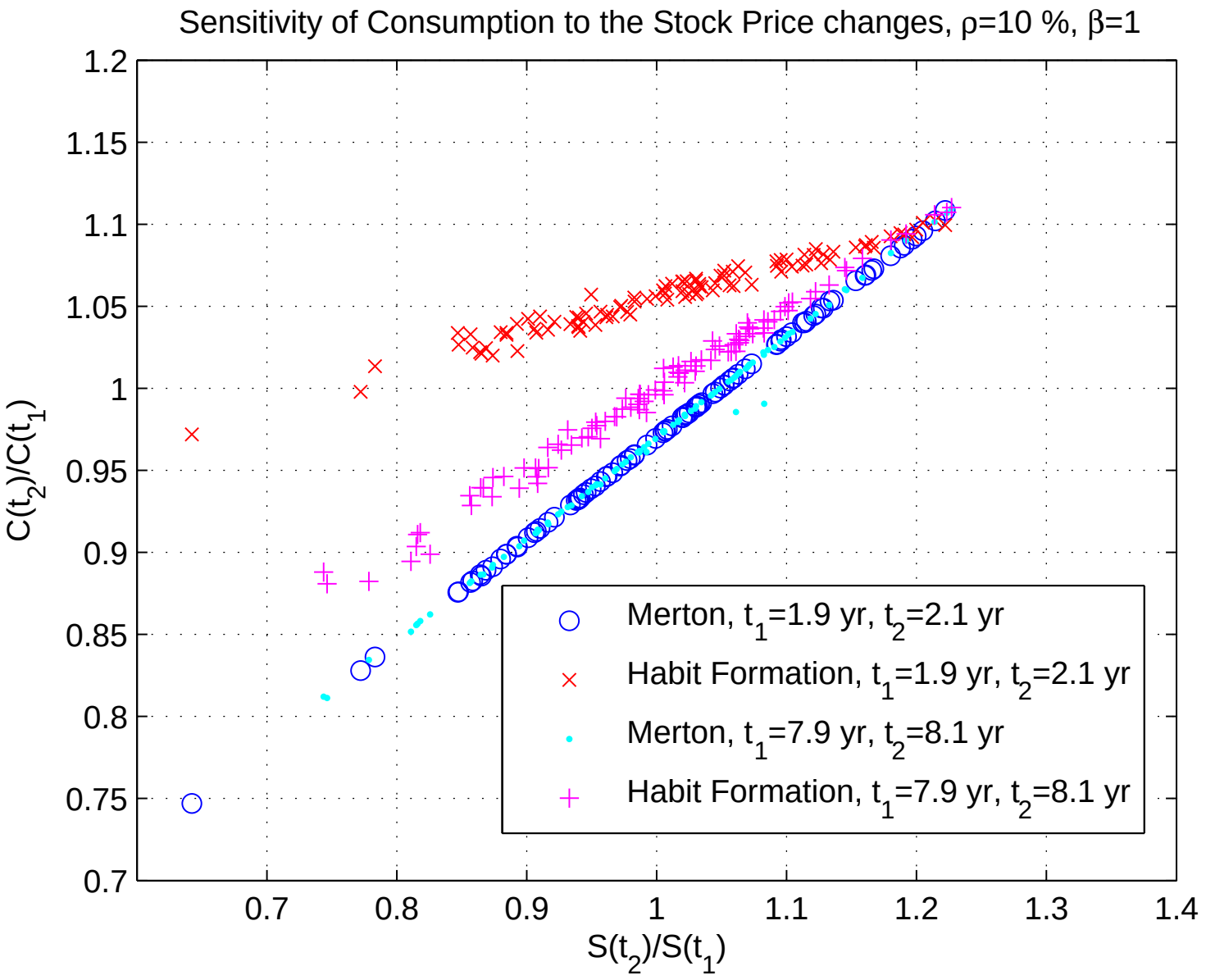

**Figure 9**. The sensitivity of consumption to the market fluctuations

As memory parameter increases, an investor intertemporally becomes more wealthy, as we see from the Fig. 10, since the Habit Formation makes him consume less in the beginning.

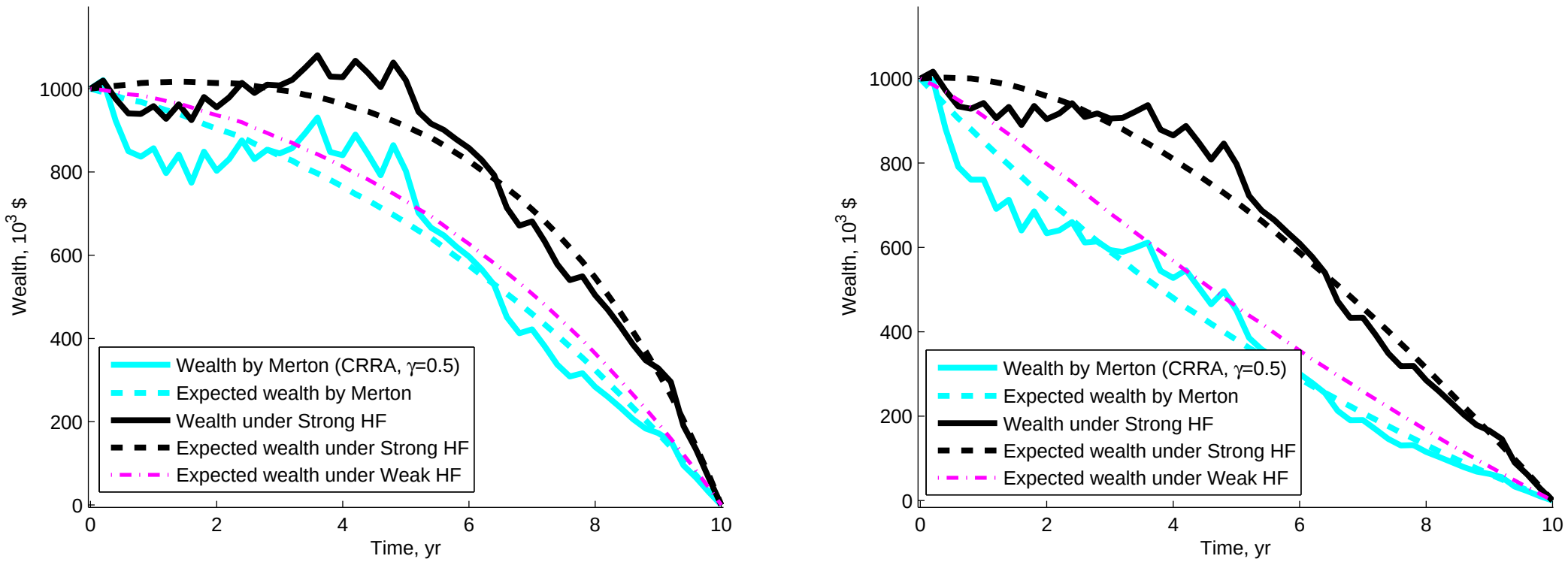

**Figure 10**. Wealth under Strong Habit Formation $\beta = 1.0$ with $\rho = 0\%$ and $\rho = 10\%$

The wealth dependence of the optimum policies is shown on Figs. 11 and 12.

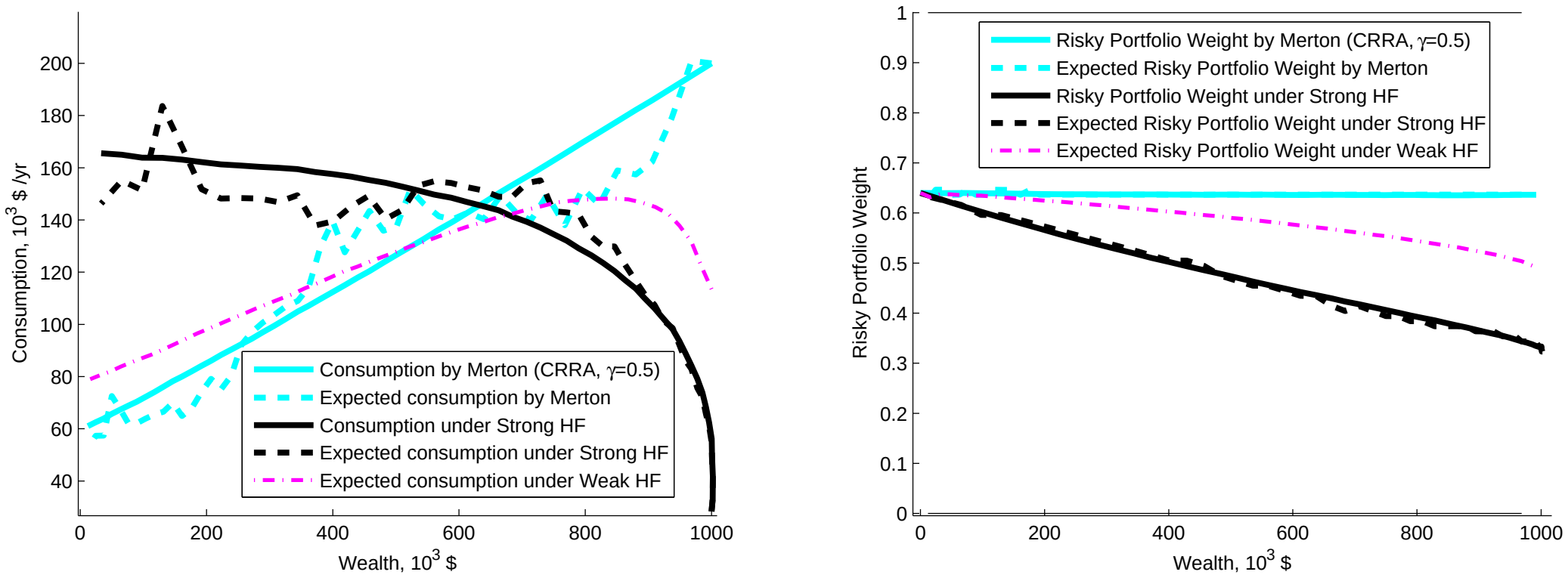

**Figure 11**. Consumption and risky portfolio weight under Strong Habit Formation $\beta = 1.0$ with $\rho = 10\%$

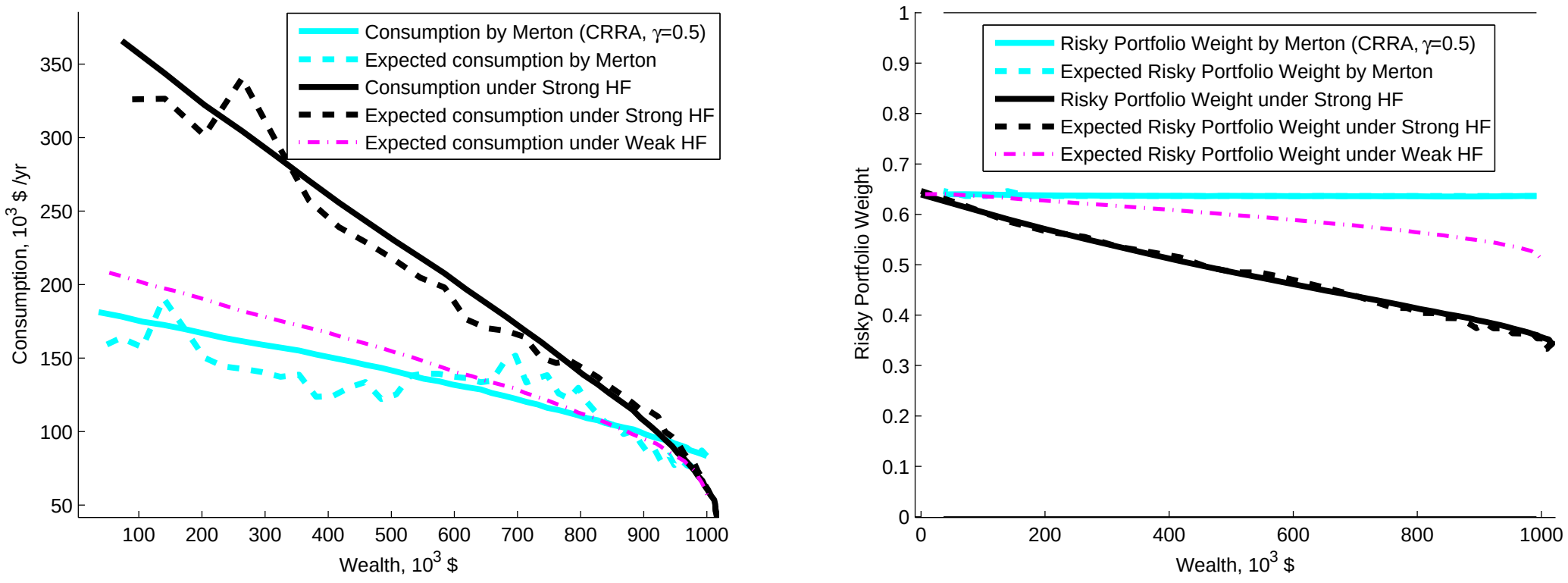

**Figure 12**. Consumption and risky portfolio weight under Strong Habit Formation $\beta = 1.0$ with $\rho = 0\%$

As $\beta$ increases, the slope of the optimum consumption as a function of wealth decreases and/or becomes more and more negative showing once again that the strong habit persistence of an investor makes him to consume less in the beginning when he has large wealth.

It is worth noting that on the graphs above the dependence on wealth is mixed with the time dependence, since the wealth itself depends on time, and it is not clear which one dominates over the other. To avoid this, in the next subsection we consider a bequest function which makes the wealth stay at approximately the same level allowing us to separate the time dependence for any given wealth level.

## 5.5 Non-zero Bequest

Consider the bequest function of the form (3.3) that changes investor's preferences making him to save some money at the time horizon in contrast to consuming all the wealth in the case of zero bequest $b = 0$. We choose the bequest parameter in such a way that the expected terminal wealth of an investor is equal to the initial wealth. For our set of parameters, this requires that $b = 0.39$ for $\rho = 0\%$ and $b = 0.62$ for $\rho = 10\%$. Now, the wealth maintains approximately the same value $W(t) \approx W_0 = \$1,000,000$ (see Fig. 13).

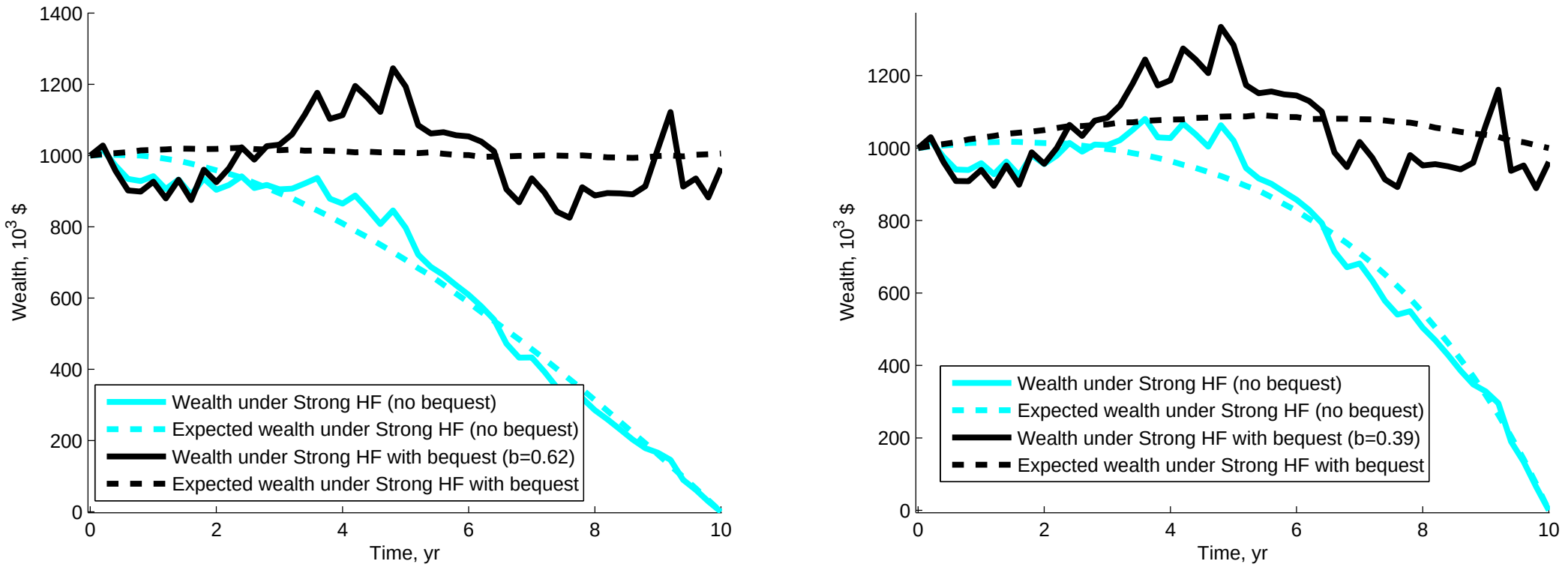

**Figure 13**. Wealth under Strong Habit Formation $\beta = 1.0$ with and without bequest, $\rho = 10\%$ (left) and $\rho = 0\%$ (right)

We see that the expected wealth for the trivial discount factor $\rho = 0\%$ has a convex shape, while the expected wealth for $\rho = 10\%$ is almost linear. Another interesting feature of a bequest function is that stock price fluctuations have much bigger impact on wealth comparing to the no-bequest case, since in the former case we don't have a "boundary condition" making the terminal wealth zero for any particular realization. With a bequest, we have a "free" boundary for wealth at the time horizon.

Since an investor's wealth remains at approximately the same level in the bequest case, he must consume with much less rate comparing to the zero bequest case, on average consuming only the return of his investments. Fig. 14 and 15 compare consumption and investment policies under Strong Habit Formation with and without bequest.

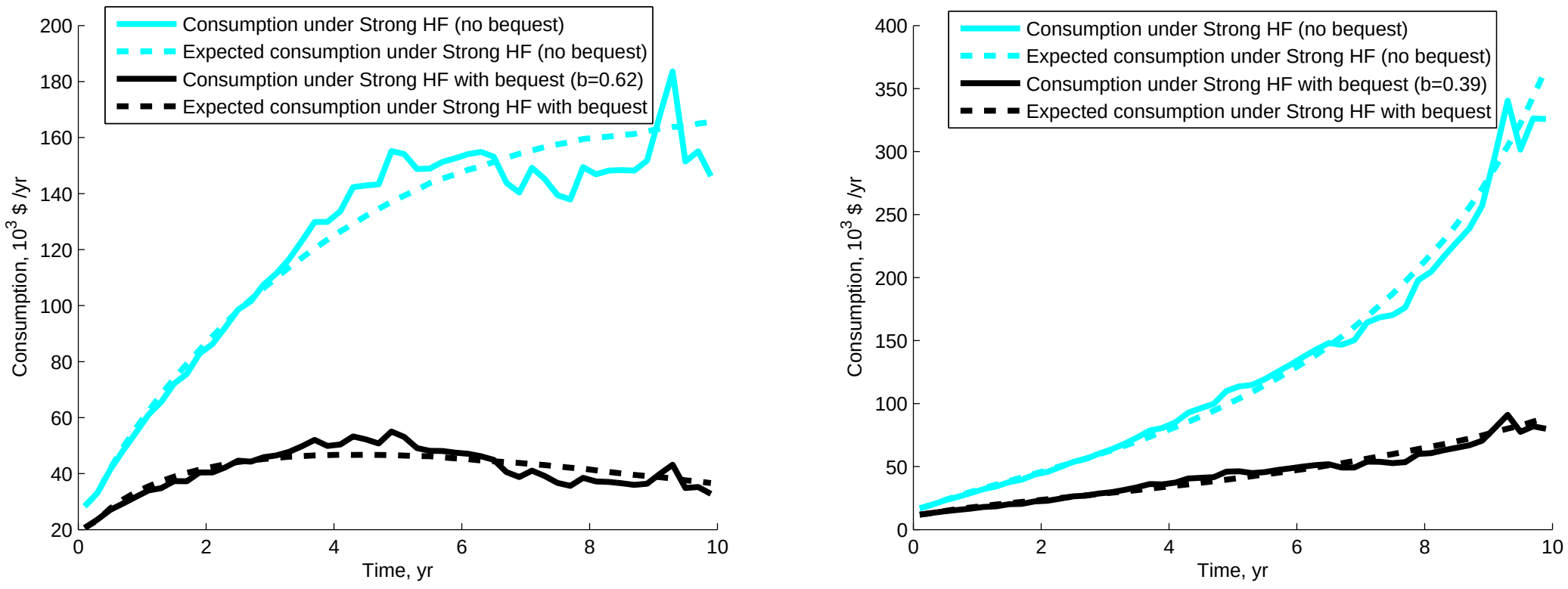

**Figure 14**. Consumption under Strong Habit Formation $\beta = 1.0$ with and without bequest, $\rho = 10\%$ (left) and $\rho = 0\%$ (right)

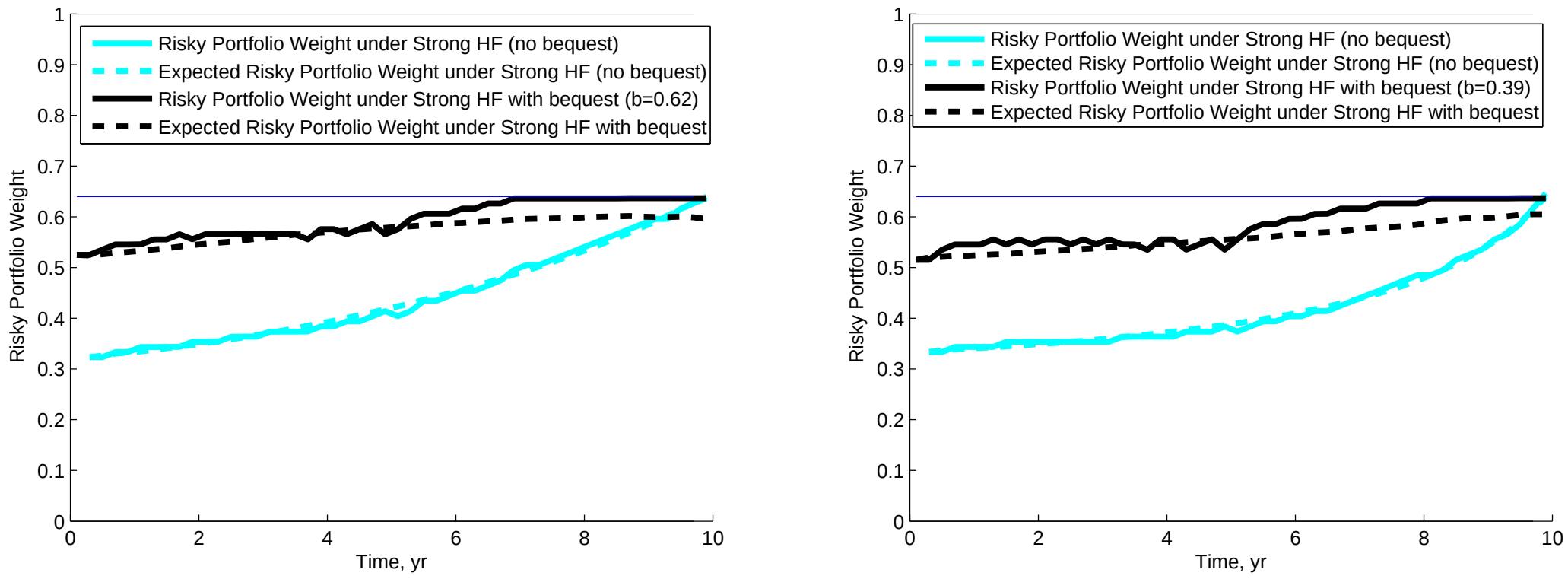

**Figure 15**. Risky portfolio weight under Strong Habit Formation $\beta = 1.0$ with and without bequest, $\rho = 10\%$ (left) and $\rho = 0\%$ (right)

Because of a "free" boundary condition at the time horizon in the bequest case, the portfolio weight feels the fluctuations of the stock price and is anti-correlated to it. Therefore, as the stock price goes up, the investor's wealth increases, and he becomes more risk averse.

Again, we plot the dependencies of consumption and portfolio weight against the expected wealth (see Figs. 16 and 17).

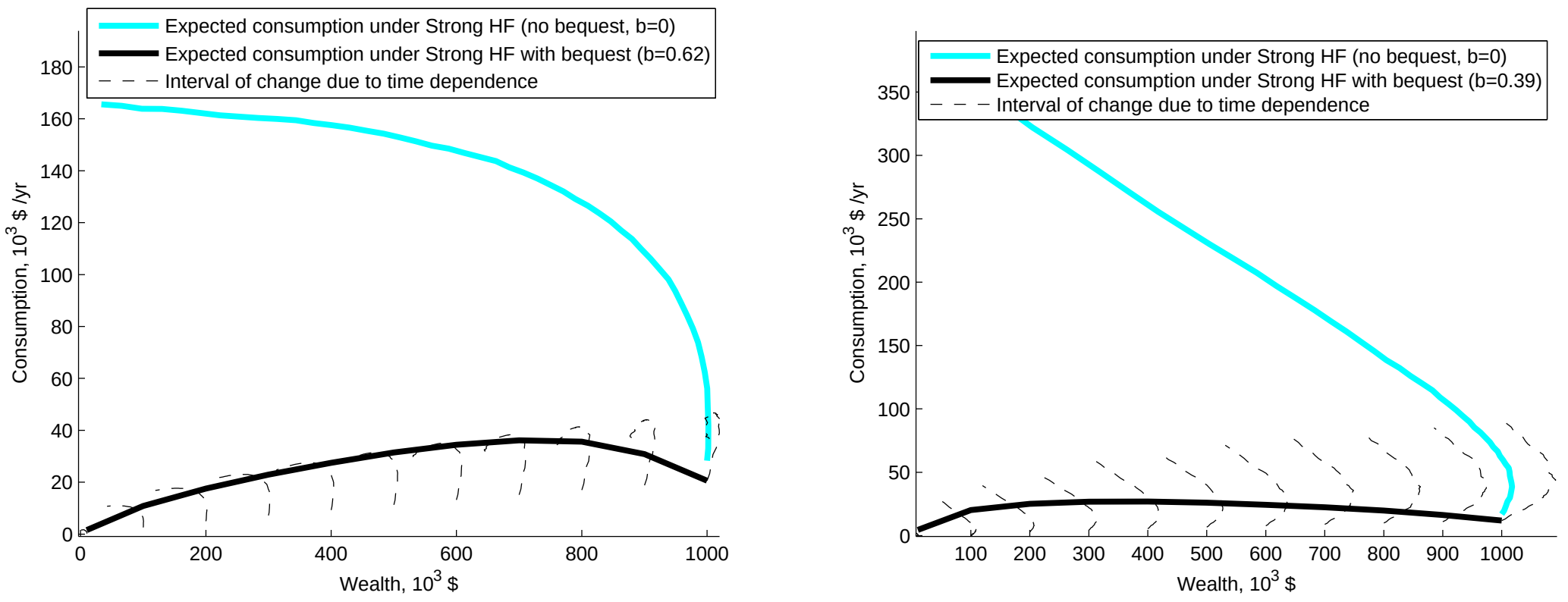

**Figure 16**. Consumption under Strong Habit Formation $\beta = 1.0$ with and without bequest, $\rho = 10\%$ (left) and $\rho = 0\%$ (right)

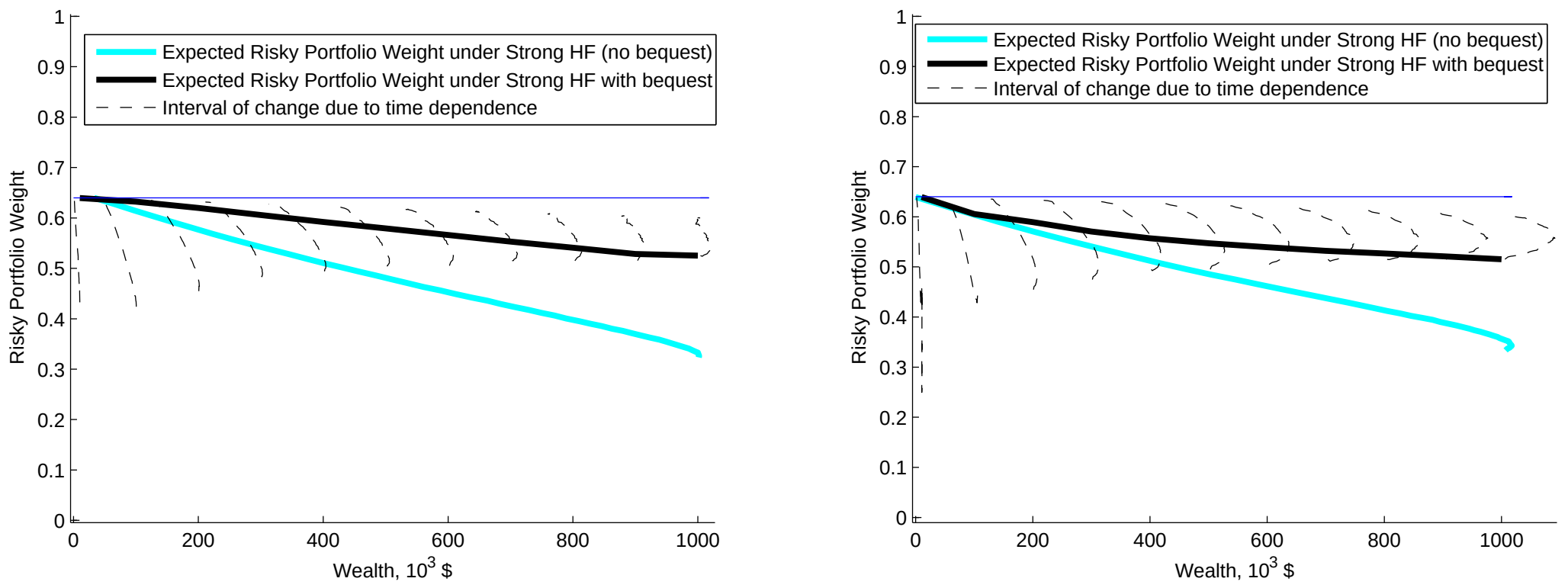

**Figure 17**. Risky portfolio weight under Strong Habit Formation $\beta = 1.0$ with and without bequest, $\rho = 10\%$ (left) and $\rho = 0\%$ (right)

We see that in the case of non-zero bequest function, the dependence of both consumption and portfolio weight on wealth is much more weak. This is the consequence of the fact that the investor has to maintain the level of his wealth and therefore has less “freedom” in choosing his consumption and portfolio.

## 5.6 Discussion

As we mentioned earlier, some of the results presented here challenge accepted financial intuition. For our habit formation specification, an investor’s risk aversion decreases with age and increases with wealth in apparent contrast to the empirical evidence. This partly can be explained by the fact that the proposed new theory doesn't model non-market risks which are always present among investor's considerations. More precisely, our model (as well as Merton's) is for a person who

- has no job;
- consumes assets with zero transaction costs;
- certain about his time horizon, health etc;

Under such limitations, according to (3.2), increasing consumption makes us happier today, but decreases the utility of future consumption. Therefore, habit formation penalizes consumption increase relative to the Merton case, making the utility function of an investor more risk averse: the more time left, the more the increase in risk aversion is, which agrees with our numerical results. This qualitative analysis shows that mathematically the model behaves correctly, but it is not a good description of a real investor, whose description requires improvements in the model to account for the bullet points shown above.

Therefore, the results should be considered as prescriptive rather than descriptive and the main attention should be paid to the differences from the Merton results, which however may suggest some qualitative changes in the optimum consumption/investment policies of an investor who bases his decisions on the standard Merton's model.

# 6 Conclusions

The present paper introduces a new model of habit formation into the consumption/investment optimization problem. The utility function explicitly depends on the averaged past consumption $\bar{C}$ and is taken in the form $U[C,\bar{C}] = U[C/(C_0 + \beta\bar{C})]$. The model has an advantage over the "popular" additive utility models (see e.g. [17]) by relaxing some constraints on the consumption path and providing optimum solutions that don't lead to bankruptcy before the time horizon.

The drawback of our model is that it doesn't allow for an explicit analytic solution, however the general properties of the solution can be easily investigated numerically. The numerical results in the paper are obtained using the powerful method of stochastic dynamic program-ming. The algorithm can operate with an arbitrary utility function and is not limited only to the HARA-type utility functions. The calculations were done by coding the algorithm in C++ and the graphs were created with the help of MATLAB.

The consumption and the portfolio weight optimum policies show qualitative differences from the Merton's solution as well as from the addictive-type habit formation models. We believe that the issue of non-addictive habit formation has not yet been properly addressed in the literature and we hope that the present paper containing explicit examples and numerical results will help to build some new intuition about the investor's consumption/investment optimization problem under habit formation.

## Acknowledgements

This work was made possible by the facilities of the Shared Hierarchical Academic Research Computing Network (SHARCNET: www.sharcnet.ca). R.N. would like to thank Yosua J. Setyobudhi for fruitful discussions on the numerical algorithms.